\documentclass[12pt,a4paper,onecolumn]{iopart}
\usepackage{graphics,graphicx}
\usepackage[small]{caption}
\newcommand{\mgb}{MgB$_2$}
\begin{document}
\title[Anomalous hysteresis in the mw surface resistance of MgB$_2$]{Anomalous magnetic hysteresis in the microwave surface resistance of \mgb\ superconductor}

\author{A. Agliolo Gallitto, M. Bonura and M. Li Vigni}

\address{CNISM and Dipartimento di Scienze Fisiche ed
Astronomiche, Universit\`{a} di Palermo,\\ Via Archirafi 36,
I-90123 Palermo, Italy}

\ead{marco.bonura@fisica.unipa.it}

\begin{abstract}
We report experimental results of the field-induced variations of
the microwave surface resistance in samples of \mgb, produced by
different methods. By sweeping the DC magnetic field at increasing
and decreasing values, we have detected a magnetic hysteresis that
can be ascribed to the different magnetic induction, due to the
critical state of the fluxon lattice. The hysteresis observed in
the bulk samples has an unusual shape, which cannot be justified
in the framework of the critical-state models.\\
\\
\noindent\emph{Keywords}: magnesium diboride; microwave response;
irreversibility properties.
\end{abstract}

\section{Introduction}
The $H$-$T$ phase diagram of type-II superconductors (SC) is
characterized by the presence of the irreversibility line,
$H_{irr}(T)$, below which the magnetic properties of the SC become
irreversible~\cite{yeshurun}. The application of a DC magnetic
field, $H_0$, smaller than $H_{irr}(T)$ develops a critical state
of the fluxon lattice, characterized by the critical current
density, $J_c$~\cite{BEAN,KIM}. The main consequence of the
critical state is the hysteretic behavior of the magnetization
curve, which gives rise to hysteresis in all of the properties
involving the presence of fluxons.

Fluxon dynamics can be conveniently investigated by measuring the
microwave (mw) surface resistance, $R_s$, which is proportional to
the mw energy losses~\cite{golo,dulcicvecchio}. Indeed, the
variations of $R_s$, induced by magnetic fields higher than the
first penetration field, are due to the presence and motion of
fluxons within the mw-field penetration depth. Most models for the
field-induced mw losses~\cite{golo,dulcicvecchio,CC,brandt} assume
an uniform distribution of fluxons in the sample, disregarding the
critical-state effects. Very recently, we have investigated, both
experimentally and theoretically, the field-induced variations of
$R_s$ in SC in the critical state~\cite{noistatocritico} and have
accounted for the magnetic hysteresis in the $R_s(H_0)$ curves of
Nb samples, powder and bulk~\cite{noiisteresi}.

Here, we report on the magnetic-field-induced variations of $R_s$
in samples of \mgb\ (bulk and powder), produced by different
methods. In the $R_s(H_0)$ curve, we have detected a magnetic
hysteresis that could be due to the different values of the
magnetic induction at increasing and decreasing field when the
sample is in the critical state. The results obtained in powder
sample can by justified in the framework of the critical state
models, while those obtained in the bulk samples show a magnetic
hysteresis of anomalous shape.

\section{Experimental Apparatus and Samples}\label{apparatus}
The field-induced variations of $R_s$ have been studied in four
samples of \mgb, three bulk samples (labelled as B) and a powdered
one (P). Samples B1 and B2 have been prepared by the one-step
method~\cite{putti}; in particular B1 has been obtained using
$^{11}$B and B2 using $^{10}$B. Sample B3 has been produced by the
reactive liquid Mg infiltration in $^{10}$B
powder~\cite{giun-cryo06}. Sample P consists of 5~mg of
Alpha-Aesar \mgb\ powder, with grain mean diameter $\approx
100~\mu$m.

\begin{figure}[h]
\centering
\includegraphics[width=18pc]{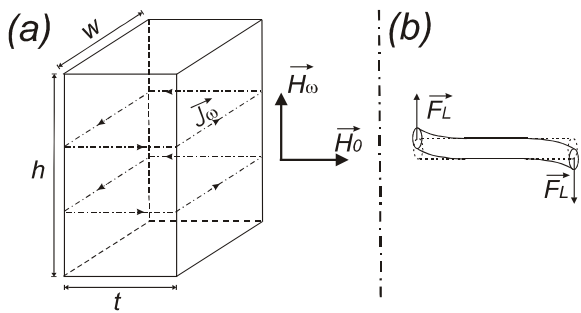}\hspace{2pc}%
\begin{minipage}[b]{16pc}\caption{\label{sample}(a)
Field and current geometry at the surface of the sample. (b)
Schematic representation of the motion of a flux line.}
\end{minipage}
\end{figure}

The mw surface resistance is measured by the cavity-perturbation
technique~\cite{golo}. A copper cavity, of cylindrical shape with
golden-plated walls, is tuned in the TE$_{011}$ mode resonating at
$\omega/2\pi\approx$~9.6 GHz. The sample is located inside the
cavity, in the region in which the mw magnetic field is maximum.
The cavity is placed between the poles of an electromagnet, which
generates DC magnetic fields up to $\mu_0H_0\approx~1~\mathrm{T}$.
Two additional coils, independently fed, allow compensating the
residual magnetic field. The sample and the field geometry is
shown in Fig.~1(a). In the mixed state, the induced mw current
causes a tilt motion of the whole vortex lattice~\cite{brandt};
Fig.~\ref{sample}(b) schematically shows the motion of a flux
line, induced by the Lorentz force, $F_L$. The mw surface
resistance of the sample is determined measuring the variation of
the quality factor of the cavity, induced by the sample, using an
hp-8719D Network Analyzer.

\section{Experimental Results}
The field-induced variations of $R_s$ have been investigated at
fixed temperatures, on increasing and decreasing the DC magnetic
field $H_0$. In all of the investigated samples, we have observed
a magnetic hysteresis in a wide range of temperatures, up to few K
below $T_c$. However, in this paper we report the results obtained
at low temperatures, where the critical-state effects are more
significant. All measurements have been performed in
zero-field-cooled samples.

\begin{figure}
\centering
\includegraphics[width=0.8\textwidth]{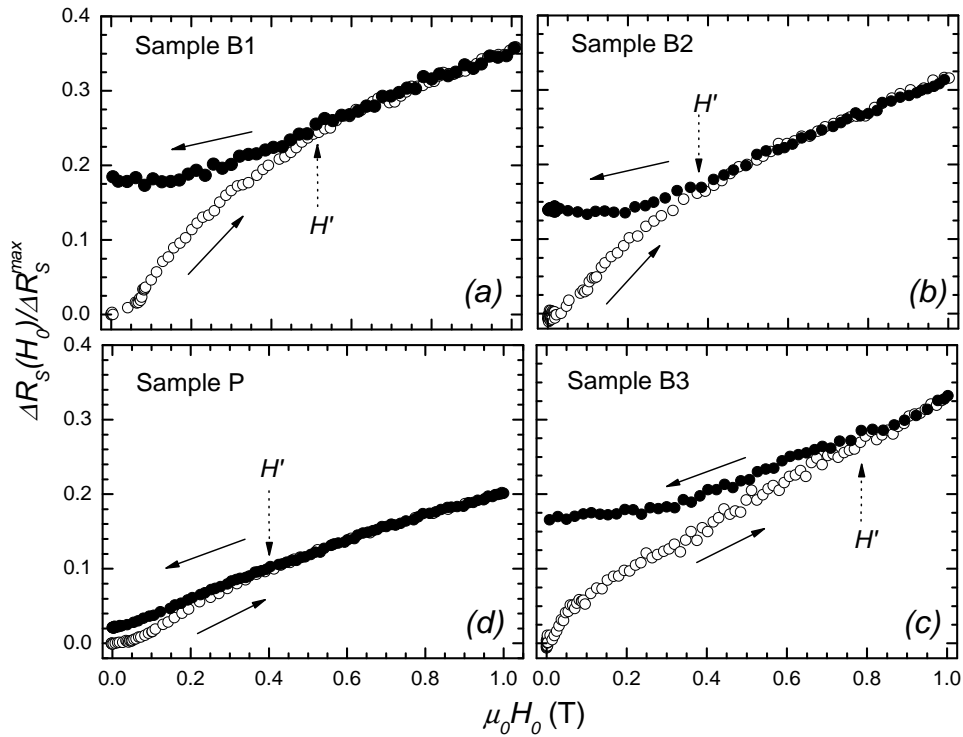}
\caption{\label{insieme}Field-induced variations of $R_s$, at $T=4.2$~K,
for the four samples; $\Delta R_s(H_0)\equiv R_s(H_0)-R_{res}$, where
$R_{res}$ is the residual mw surface resistance at $H_{0}=0$; $\Delta
R_s^{max}\equiv R_{n}-R_{res}$, where $R_n$ is the surface resistance in
the normal state.} \vspace{-5mm}
\end{figure}

Fig.~\ref{insieme} shows the field-induced variations of $R_s$,
obtained at $T=4.2$~K by sweeping $H_0$ from 0 up to 1~T and back,
for the four samples. $\Delta R_s(H_0)\equiv R_s(H_0)-R_{res}$,
where $R_{res}$ is the residual mw surface resistance at
$H_{0}=0$; the data are normalized to the maximum variation,
$\Delta R_s^{max}\equiv R_{n}-R_{res}$, where $R_n$ is the surface
resistance in the normal state, at $T\approx 40$~K. Open symbols
refer to the results obtained at increasing fields, full symbols
those at decreasing fields. On decreasing $H_0$, after it had
reached 1~T, we observe an initial reversible behavior followed by
a hysteretic behavior below a certain value of $H_0$, depending on
the sample, indicated in the figure as $H^{\prime}$.

In all of the investigated samples, one can observe that the
application of a DC magnetic field of $\approx 1$ T, much smaller
than the upper critical field, induces an unusually enhanced $R_s$
variation; it ranges from $\approx 20 \%$ to $\approx 40 \%$ of
the maximum variation (achieved when the sample goes to the normal
state), dependently on the sample. The enhanced $R_s$ variation
has been ascribed to the unusual fluxon structure in \mgb\ SC, due
to the two superconducting gaps~\cite{shibata,noiPRB}; in this
paper, we would discuss the irreversible properties of $R_s$ in
\mgb.

A comparison among the results in the different samples, reported
in Fig.~\ref{insieme}, shows that in all of the bulk samples the
width of the hysteresis loop at $H_0=0$ is $\approx 50 \%$ of the
$R_s$ variation observed at $\mu_0H_0=1$ T, while in the powder
sample it is $\approx 10 \%$. Another peculiarity distinguishing
the mw responses of the powder and the bulk samples is the shape
of the decreasing-field branch of the $R_s(H_0)$ curve. Indeed,
one can see that in the powder sample the decreasing-field branch
exhibits a monotonic decrease, starting from $H_0=H^{\prime}$ down
to $H_0=0$. On the contrary, in the bulk samples the
decreasing-field branch of the $R_s(H_0)$ curve shows an
unexpected plateau extending from $\mu_0 H_0\approx 0.2$ T down to
0.

\begin{figure}[h]
\centering
\begin{minipage}{17pc}\centering
\includegraphics[width=16pc]{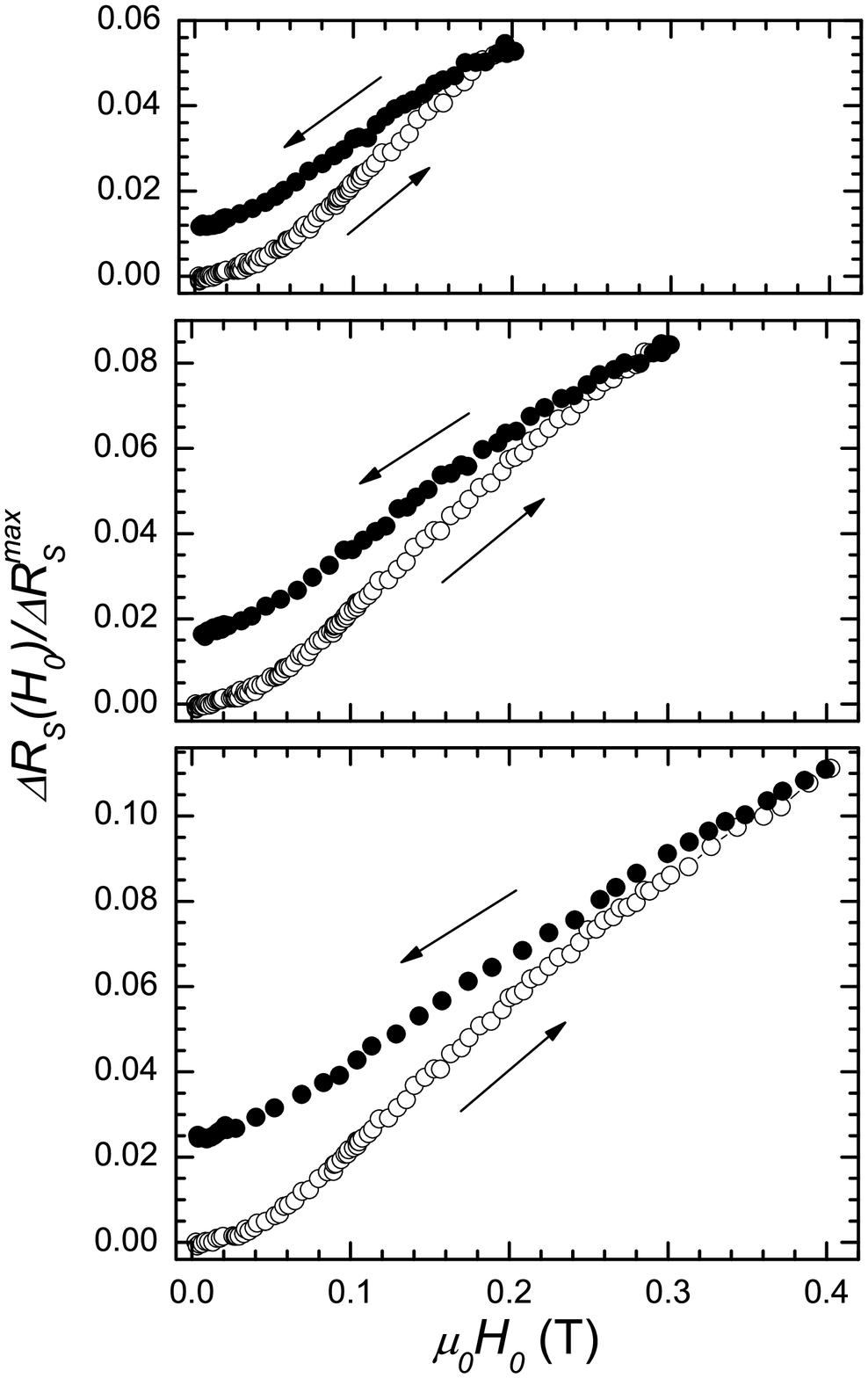}
\caption{\label{polvere}$R_s(H_0)$ curves, obtained for sample P
by sweeping $H_0$ up to a certain value, $H_{max}$, and back, for
different values of $H_{max}$; $T=4.2$~K.}
\end{minipage}\hspace{2pc}%
\begin{minipage}{17pc}\centering
\includegraphics[width=16pc]{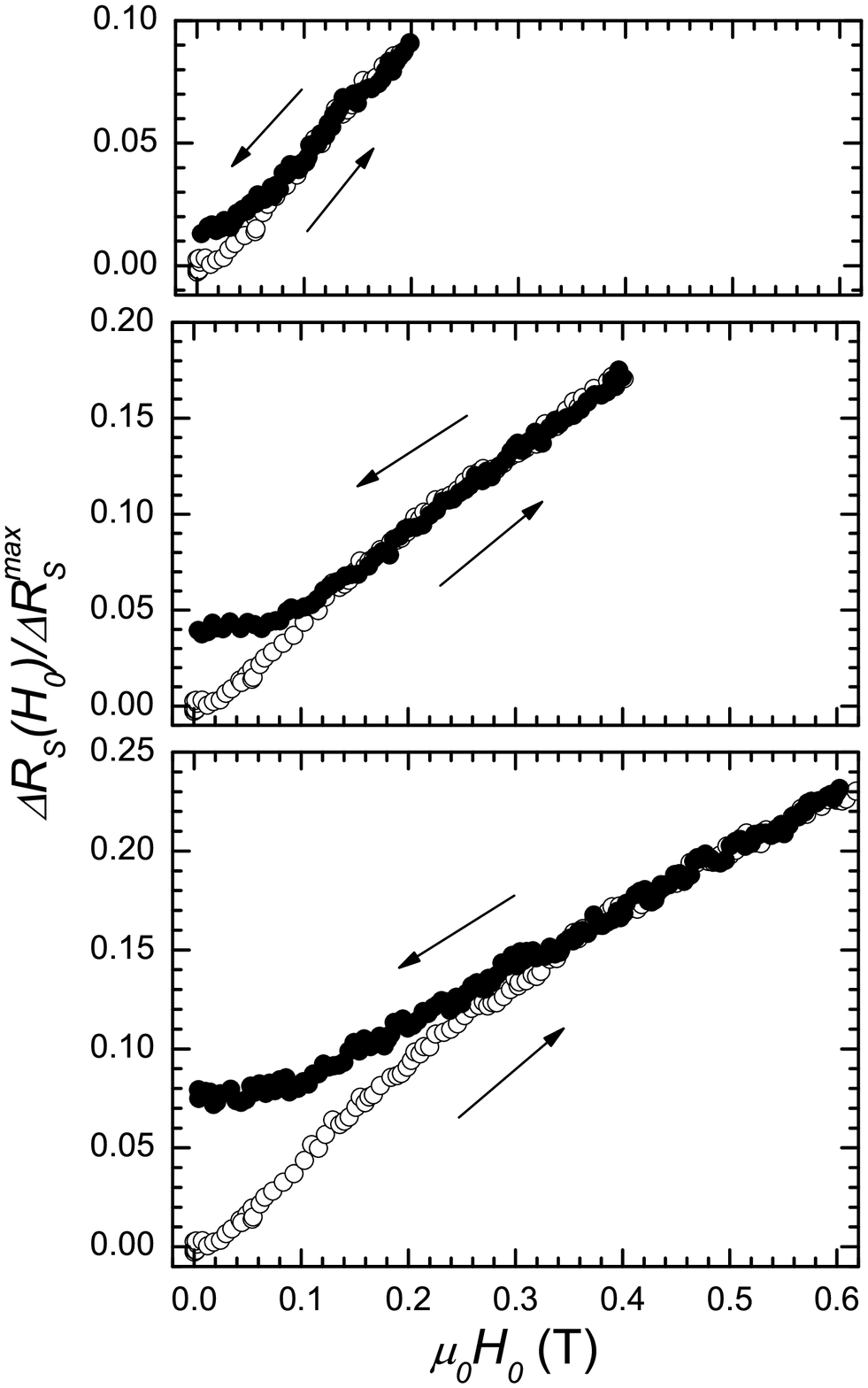}
\caption{\label{bulk}$R_s(H_0)$ curves, obtained for sample B2 by
sweeping $H_0$ up to a certain value, $H_{max}$, and back, for
different values of $H_{max}$; $T=4.2$~K.}
\end{minipage}\vspace{-5mm}
\end{figure}

The field-induced variations of $R_s$ have also been investigated
by cycling the DC magnetic field in different ranges.
Figs.~\ref{polvere} and \ref{bulk} show the $R_s(H_0)$ curves,
obtained for samples P and B2, respectively, by sweeping $H_0$
from zero up to a certain value, $H_{max}$, and back, for
different values of $H_{max}$. As expected, due to the different
trapped flux, the smaller $H_{max}$ the smaller the hysteresis
width. In sample P, when $H_{max}$ is smaller than the
$H^{\prime}$ value of Fig.~\ref{insieme}(d), the $R_s(H_0)$ curve
is irreversible in the whole field range swept; we remark that we
have not reported results obtained with $H_{max}> H^{\prime}$
because they exactly reproduce those of Fig.~\ref{insieme}(d). On
the contrary, in the bulk sample the value of $H_0$ below which
the decreasing-field branch deviates from the increasing-field one
depends on $H_{max}$. For $\mu_0H_{max}=0.2$~T, the hysteresis is
visible in a restricted range of fields and the decreasing-field
branch shows a monotonic decrease; instead, for $\mu_0 H_{max}>
0.2$~T the unexpected plateau at low magnetic fields is well
visible. We would remark that similar results have been obtained
in all of the bulk \mgb\ samples investigated.

\section{Discussion}
As it is well known, the magnetic-field-induced variations of the
mw surface resistance are due to the presence and motion of
fluxons within the mw-field penetration
depth~\cite{golo,dulcicvecchio,CC,brandt}; in particular, at low
temperatures and for applied magnetic fields lower enough than the
upper critical field, they are essentially due to vortex motion
induced by the mw current. In most of the models for fluxon
dynamics reported in the literature, the fluxon distribution is
assumed uniform, neglecting the effects of the critical state of
the fluxon lattice. On the other hand, a magnetic hysteresis in
the $R_s(H_0)$ curve is expected in SC in the critical state
because of the different fluxon density at increasing and
decreasing DC fields. In order to account for this hysteretic
behavior, it is essential to consider the fluxon distribution in
the sample determined by the critical current density $J_c$.

Recently, we have investigated the field dependence of the mw
surface resistance of SC in the critical
state~\cite{noistatocritico,noiisteresi}, by taking into due
account the fluxon distribution, and we have quantitatively
justified the hysteretic behavior of the $R_s(H_0)$ curve detected
in Nb samples~\cite{noiisteresi}. The field geometry we have used
(see Fig.~\ref{sample}) is particularly convenient to investigate
such effects essentially for two reasons. Firstly, the effects of
the non-uniform fluxon distribution on $R_s$ are enhanced because
in the sample surfaces normal to the external magnetic field the
mw current, penetrating along the fluxon axis within the mw-field
penetration depth, bends the end segments of all the fluxons.
Moreover, just in this case, one can calculate the average value
of $R_s$ by integration over the sample surfaces that contribute
to the mw energy losses. We have shown that the parameter that
mainly determines the features of the $R_s(H_0)$ curve is the full
penetration field, $H^*$. In particular, the hysteresis width is
related to the value of $H^*$; samples of small size and/or small
$J_c$ are expected to exhibit weak hysteretic behavior. $H^*$
determines also the shape of the hysteresis loop. On increasing
$H_0$ from 0 up to $H^*$, more and more sample regions contribute
to the mw losses, giving rise to a positive curvature of the
increasing-field branch of the $R_s(H_0)$ curve. For $H_0 > H^*$,
in the whole sample the local magnetic induction depends about
linearly on the external magnetic field and the increasing-field
branch is expected to have a negative concavity. The shape of the
decreasing-field branch is strictly related to the shape of the
magnetization curve; it should exhibit a negative concavity, with
a monotonic reduction of $R_s$ in the whole field range swept.

A comparison among the results reported in the different panels of
Fig.~\ref{insieme} shows that in the powder sample the hysteresis
width is smaller than that observed in the bulk samples. In the
framework of the critical-state model, this finding is ascribable
to the small value of $H^*$ due to the reduced size of the powder
grains, with respect to the bulk samples. Furthermore, the
decreasing-field branch of the $R_s(H_0)$ curve of sample P
exhibits a monotonic decrease, starting from $H_0=H^{\prime}$ down
to $H_0=0$, as expected. On the contrary, in all of the bulk
samples the shape of the hysteresis shows several anomalies. The
most unexpected behavior concerns the plateau observed in the
decreasing-field branch of the $R_s(H_0)$ curve, after the sample
had been exposed to relatively high fields. The presence of this
plateau is puzzling because it would suggest that the trapped flux
does not change anymore on decreasing $H_0$ below a certain
threshold value, of the order of 0.1~T. A further anomaly is
visible in the increasing-field branch, which exhibits negative
concavity in the whole field range swept, despite the estimate
value of $H^*$ for the bulk samples is few Tesla~\cite{noiPRB}. On
the contrary, in the $R_s(H_0)$ curve of sample P one can note a
change of concavity in the increasing-field branch at
$\mu_0H_0\sim 0.1$~T, which is a reasonable value of $H^*$ for the
powder sample.

We would remark that the magnetic field value at which the
decreasing-field branch of the $R_s(H_0)$ curve deviates from the
increasing-field one could differ from $H_{irr}(T)$ deduced from
magnetization measurements. Indeed, it has been shown that, in
samples of finite dimensions, the application of AC magnetic
fields normal to the DC field may induce the fluxon lattice to
relax toward an uniform flux distribution~\cite{brandt3}, reducing
the value of $H_{irr}(T)$. The process is particularly relevant
for thin samples and/or small critical current. Furthermore,
considering the sensitivity of our experimental apparatus, we
expect to detect hysteresis for $J_c$ greater than $\sim
10^4~\mathrm{A/cm}^2$. In this framework, one can justify the
reduced value of $H^{\prime}$ we obtained in sample P (see
Fig.~\ref{insieme}(d)), which is roughly one order of magnitude
smaller than the value estimated using $J_c$ values reported in
the literature for \mgb.

The same justification cannot be given for the results obtained in
the bulk samples. From Fig.~\ref{bulk} one can see that, after the
inversion of the field-sweep direction, the $R_s(H_0)$ curve shows
an initial reversible behavior, independently of the value of
$H_{max}$. As a result, the value of the magnetic field at which
the decreasing-field branch deviates from the increasing-field one
depends on $H_{max}$. We would remark that this result has been
obtained in all of the bulk samples we have investigated. This
finding disagrees with the meaning of the irreversibility line.
Indeed, the region of the $H$-$T$ plane in which SC exhibit
irreversible magnetic properties is not expected to depend on the
magnetic history of the sample, but only on the $J_c$ value at the
corresponding temperature and external magnetic field. On the
other hand, this anomaly has not been observed in the powder
sample; indeed, Fig.~\ref{polvere} shows that when the
magnetic-field-sweep direction is reversed at fields smaller than
the value of $H^{\prime}$ deduced from Fig.~\ref{insieme}(d) the
$R_s(H_0)$ curve is irreversible in the whole field range swept,
consistently with the expected behavior.

In the framework of the critical state, the main property
distinguishing the magnetic response of samples having small or
large dimensions is the residual magnetic induction. Let us
consider two samples with the same value of $J_c$ and different
sizes, small and large; furthermore, let us indicate as
$H_{small}^*$ and $H_{large}^*$ the full penetration field of the
small- and large-size sample, respectively. In the critical state
\`{a} la Bean, after cycling $H_0$ in the range $0\rightarrow
H_{max}\rightarrow 0$, with $H_{small}^*< H_{max}<H_{large}^*$,
the maximum value of the local magnetic induction in the samples
will be $B_{max}=0.5~\mu_0H_{small}^*$ in the small-size sample
and $B_{max}=0.5~\mu_0H_{max}$ in the large-size sample. This
means that in sample regions far from the edges, the local
magnetic field in the large-size sample can be much higher than
that in the small-size one. Since the dimensions of our bulk
samples are a factor of $\approx 20$ greater than the grain
dimensions of sample P, one may infer that the different behavior
of the irreversible properties of $R_s$ in the powder and bulk
samples is related to the higher value of the local magnetic
induction inside the bulk samples, far from the edges, when the
applied field is reduced after it had reached relatively high
values.

\section{Conclusion}
We have reported on the irreversible properties of the
field-induced variations of the mw surface resistance of different
ceramic \mgb\ samples, powder and bulk, produced by different
techniques. The results have qualitatively been discussed in the
framework of models reported in the literature. The hysteretic
magnetic behavior of $R_s$ observed in the bulk samples exhibits
several anomalies not yet understood. In particular, the magnetic
field at which the decreasing-field branch deviates from the
increasing-field one depends on the maximum magnetic field
reached; furthermore, the decreasing-field branch of the
$R_s(H_0)$ curve shows an unexpected plateau extending from a
certain value of $H_0$, depending on $H_{max}$, down to zero. In
the powder sample, consisting of \mgb\ grains of small dimensions,
the shape of the observed magnetic hysteresis is consistent with
that expected using the critical-state model. This finding
suggests that the anomalous response of the bulk samples is
strictly related to the higher value of the local magnetic field
in the interior of the sample, when the applied magnetic field is
reduced after it had reached high values. \ack
The authors thank G. Giunchi and P. Manfrinetti for having kindly
supplied the bulk \mgb\ samples.



\section*{References}

\begin{thebibliography}{99}

\bibitem{yeshurun} Y. Yeshurun, A. P. Malozemoff and A. Shaulov 1996 \emph{Rev. Mod. Phys.}
{\bf 68} 911, and Refs. therein.

\bibitem{BEAN}C. P. Bean 1962 \emph{Phys. Rev. Lett.} {\bf 8} 250.

\bibitem{KIM}Y. B. Kim, C. F. Hempstead and A. R. Strnad 1962 \emph{Phys. Rev.
Lett.} {\bf 9} 306.

\bibitem{golo}M. Golosovsky, M. Tsindlekht and D. Davidov  1996 \emph{Supercond. Sci. Technol.}
\textbf{9} 1, and Refs. therein.

\bibitem{dulcicvecchio}  A. Dul$\check{\mathrm{c}}$i$\acute{\mathrm{c}}$  and
M. Po$\check{\mathrm{z}}$ek 1993 \emph{Physica } C \textbf{218}
449.

\bibitem{CC}M. W. Coffey and J. R. Clem 1992 \emph{Phys. Rev.} B {\bf 45} 9872;
{\bf 45} 10527.

\bibitem{brandt} E. H. Brandt 1991 \emph{Phys. Rev. Lett.} {\bf 67} 2219; 1995
\emph{Rep. Prog. Phys.} \textbf{58} 1455.

\bibitem{noistatocritico} M. Bonura, E. Di Gennaro, A. Agliolo Gallitto and
M. Li Vigni 2006 \emph{Eur. Phys. J.} B {\bf 52} 459.

\bibitem{noiisteresi} M. Bonura, A. Agliolo Gallitto and M. Li Vigni
2006 \emph{Eur. Phys. J.} B {\bf 53} 315.

\bibitem{putti} M. Putti, V. Braccini, E. Galleani, F. Napoli, I. Pallecchi,
A. S. Siri, P. Manfrinetti and A. Palenzona 2003 \emph{Supercond.
Sci. Technol.} \textbf{16} 188.

\bibitem{giun-cryo06}G. Giunchi 2006 \emph{Advances in Cryogenic Engeneering} \textbf{52} 813.

\bibitem{shibata} A. Shibata, M. Matsumoto, K. Izawa, Y. Matsuda, S. Lee
and S. Tajima 2003 \emph{Phys. Rev.} B {\bf 68} 060501(R).

\bibitem{noiPRB}M. Bonura, A. Agliolo Gallitto, M. Li Vigni, C. Ferdeghini and C.
Tarantini 2007 \emph{e-print} arXiv:cond-mat/0709.0618.

\bibitem{brandt3}E. H. Brandt and G. P. Mikitik 2002 \emph{Phys. Rev. Lett.} {\bf 89}
027002; G. P. Mikitik and E. H. Brandt 2003 \emph{Phys. Rev.} B
{\bf67} 104511.

\end{thebibliography}
\end{document}